\documentclass[twocolumn,showpacs,preprintnumbers,amsmath,amssymb,floatfix]{revtex4}
\usepackage{graphicx}% Include figure files
\usepackage{epsfig}  % Include figure files
\usepackage{dcolumn}% Align table columns on decimal point
\usepackage{bm}% bold math

\def\issue(#1,#2,#3){{\bf #1}, #2 (#3)} % AIP format

\def\opcit(#1){ {\em op. cit.}, #1}
\def\etal {\em et al.}

\def\APP(#1,#2,#3){Acta Phys.\ Polon.\ \issue(#1,#2,#3)}
\def\ARNPS(#1,#2,#3){Ann.\ Rev.\ Nucl.\ Part.\ Sci.\ \issue(#1,#2,#3)}
\def\CPC(#1,#2,#3){Comp.\ Phys.\ Comm.\ \issue(#1,#2,#3)}
\def\CIP(#1,#2,#3){Comput.\ Phys.\ \issue(#1,#2,#3)}
\def\EPJC(#1,#2,#3){Eur.\ Phys.\ J.\ C\ \issue(#1,#2,#3)}
\def\EPJD(#1,#2,#3){Eur.\ Phys.\ J. Direct\ C\ \issue(#1,#2,#3)}
\def\IEEETNS(#1,#2,#3){IEEE Trans.\ Nucl.\ Sci.\ \issue(#1,#2,#3)}
\def\IJMP(#1,#2,#3){Int.\ J.\ Mod.\ Phys. \issue(#1,#2,#3)}
\def\JHEP(#1,#2,#3){J.\ High Energy Physics \issue(#1,#2,#3)}
\def\JPG(#1,#2,#3){J.\ Phys.\ G \issue(#1,#2,#3)}
\def\MPL(#1,#2,#3){Mod.\ Phys.\ Lett.\ \issue(#1,#2,#3)}
\def\NP(#1,#2,#3){Nucl.\ Phys.\ \issue(#1,#2,#3)}
\def\NIM(#1,#2,#3){Nucl.\ Instrum.\ Meth.\ \issue(#1,#2,#3)}
\def\PL(#1,#2,#3){Phys.\ Lett.\ \issue(#1,#2,#3)}
\def\PRD(#1,#2,#3){Phys.\ Rev.\ D \issue(#1,#2,#3)}
\def\PRL(#1,#2,#3){Phys.\ Rev.\ Lett.\ \issue(#1,#2,#3)}
\def\SJNP(#1,#2,#3){Sov.\ J. Nucl.\ Phys.\ \issue(#1,#2,#3)}
\def\ZPC(#1,#2,#3){Zeit.\ Phys.\ C \issue(#1,#2,#3)}

\def\sone {{\mathbb{S}}_1}
\def\ztwo {{\mathbb{Z}}_2}
\def\stwo {\sqrt{2}}

\def\spir {\sqrt{\pi R}}

\def\be {\begin{equation}}
\def\ee {\end{equation}}
\def\bea {\begin{eqnarray}}
\def\eea {\end{eqnarray}}

\def\bc {\begin{center}}
\def\ec {\end{center}}

\begin{document}

\preprint{CU-PHYSICS/07-2008}
%\preprint{arXiv:yymm.nnnn[hep-ph]}

\title{Universal Extra Dimension}

\author{Anirban Kundu}
\affiliation{Department of Physics, University of Calcutta,\\
92 A.P.C. Road, Kolkata 700009, India}
%\email{akphy@caluniv.ac.in}

\date{\today}

\begin{abstract}
This is a brief discussion of the following features of the Universal Extra 
Dimension (UED) model: (i) Formulation, (ii) Indirect bounds, (iii) Collider 
search and the Inverse Problem, (iv) Astrophysical bounds, and 
(v) UED with two extra dimensions.
\end{abstract}


%\keywords{Supersymmetry, R-parity violation, Neutral B mixing, CP violation}
\maketitle

\section{Introduction}
The Universal Extra Dimension (UED) model, or models as there are already
quite a few interesting variants of the minimal one, has a flat metric (like
ADD) and a small compactification radius of ${\cal O}$(TeV$^{-1}$) (like RS). 
Also, this is
the most democratic extra dimension (ED) model in the sense that all Standard
Model (SM) fields can propagate in the extra dimension, or bulk. Thus, in
essence, it is quite similar to the first-generation ED models of Kaluza
and Klein \cite{kk}. However, the motivations are quite different. 

Why UED? The cons first: this does not solve the gauge hierarchy problem. 
Both ADD and RS models lower the Planck mass, as seen on our brane, by two
very different but very elegant mechanisms. UED does nothing of that sort;
in fact, we will see that the model breathes more freely when we do not
include gravity. Issues like the stabilisation of the radius of the extra 
dimension are not addressed. If you think that the fine-tuning of the Higgs
mass is the most serious issue in particle physics, you will probably not turn
to this model.

Now the pros. First, the dark matter. The dark matter provides almost one-fourth
of the energy density of the universe, and UED, among all extra dimensional 
models, supplies a very good candidate for the cold dark matter (CDM).
In fact, it is the most theoretically motivated candidate --- {\em e.g.}, 
in supersymmetry, we impose the conservation of R-parity by hand and get
the neutralino dark matter; but the UED dark matter is a necessary consequence
of the formulation. Second, the dark matter constraint and the
indirect limits on the compactification radius guarantee a spectrum that
is completely within the reach of LHC. The first excited states of the SM 
particles should be between 400-900 GeV. 
 
As has been pointed out, the collider signals mimic those of
supersymmetry; so this is probably the most serious case of the so-called
LHC inverse problem --- the discrimination of SUSY and UED from signals.
In fact, I will discuss the issue of discriminating UED from other NP
models too, not confining only to R-parity conserving SUSY.  

UED scores more positive points if one considers two extra dimensions.
As I will touch upon later, the 6-d UED model answers two very important
questions: Why proton lifetime is so large? (This is one of the most 
challenging problems in the extra dimensional models.) Why there are
three generations?  

The plan is as follows: I will discuss, in the subsequent sections, 
(i) Formulation, (ii) Indirect bounds, (iii) Collider search prospects,
and discrimination from other models, most notably supersymmetry, 
(iv) Astrophysical bounds, and inclusion of gravity, and 
(v) UED with two extra dimensions. Except for the last part, I will
concentrate on the minimal version of UED. There are at least three other talks
that will focus on several interesting features of the model: the phenomenology
of the scalar sector of minimal UED \cite{biplob} and of 6-d UED 
\cite{kirtiman}, and the power-law evolution of the gauge couplings in
UED \cite{swarup}.

\section{Formulation}
The basic idea of Appelquist, Cheng, and Dobrescu \cite{acd} is very simple:
apart from four large dimensions $x^\mu$, there is a small dimension $y$
(this I will call the bulk),
which is compactified on a circle of radius $R$. The points $y$ and $y+2\pi R$
are identified. This means that the momentum along the fifth direction, $p_5$,
is discrete: $p_5 R = n$, where $n$ is any integer. Just like 
particle in a box, there will be equispaced states, whose masses are given by
\be
m_n^2 = m_0^2 + n^2/R^2\,,  
\label{kktree}
\ee
where the last term comes from $p_5^2$. The integer $n$ is called the
Kaluza-Klein (KK) number; $n=0$ corresponds to the zero mode. 
{\em Every SM particle is associated with an equispaced tower of particles
of identical quantum numbers.} 
As all particles can access the bulk, momentum along the fifth dimension,
and hence $n$, is conserved in any process. Thus, (i) all $n\not= 0$ 
particles are to be pair produced in collider experiments
\footnote{One may question how, in the process of one $n=0$ particle decaying 
into two $n=1$ particles, KK number is conserved. Note that while performing
the Fourier expansion of the 5-d field, we summed over all {\em positive}
integers only. Fields with negative $n$ are identical in all respect with
fields with positive $n$. So, while considering the KK-number conservation,
one should consider both choices: $|n_1\pm n_2|$. If one of the combination
fits, that's fine. The complexity arose because KK-number is an additive
quantum number and not a multiplicative one like R-parity.}; (ii) the
lightest particle for each $n$ level is stable. Note that
if $R^{-1}$ is of the order of a few hundred GeV, the particles for each
$n>0$ level are quasi-degenerate. This degeneracy is somewhat lifted by
the radiative corrections \cite{cms1,georgi}. 

In the minimal 5d UED model, the fermions are necessarily vectorial, even
in their zero modes. This
can simply be guessed from the fact that $\gamma_5$ is a part of the
$\gamma$-matrix set itself: $\Gamma^M = (\gamma^\mu,i\gamma_5)$. To get
chiral fermions in zero mode, we need a further $\ztwo$ orbifolding, so
that the compactified dimension becomes $\sone/\ztwo$. This is nothing but
a fold of the circle along one of its diameters and identification of the 
points $y$ and $-y$ for $-\pi R \leq y \leq \pi R$ (remember that there is
already the identification of $y\to y+2\pi R$). Fields can be even or odd
under this $\ztwo$: the Higgs field, the first four components of the gauge
fields, the right-chiral component of SU(2) singlet fermions, and left-chiral
components of SU(2) doublet fermions are all even, while the fifth component
of the gauge fields, the left-chiral component of SU(2) singlet fermions,
and right-chiral components of SU(2) doublet fermions are all odd.  

The 5d fields can be Fourier expanded as
\bea
\phi_+(x^\mu,y) &=& \frac{1}{\spir} \phi_+^{(0)}(x^\mu) + \frac{\stwo}{\spir}
\sum_{n=1}^\infty \cos\frac{ny}{R} \phi_+^{(n)} (x^\mu),\nonumber\\
\phi_-(x^\mu,y) &=& \frac{\stwo}{\spir}
\sum_{n=1}^\infty \sin\frac{ny}{R} \phi_-^{(n)} (x^\mu)\,,
\eea
where $\phi_+$ is even and $\phi_-$ is odd under $\ztwo$. Fields which are
odd under the $\ztwo$ orbifold symmetry do not have zero modes. Only even
fields have zero modes, which are identified with the SM particles.
There are two fixed points on the orbifold that are mapped onto 
themselves: $y=0$ and $y=\pi R$. 
The even and odd fields satisfy $\partial_5 \phi_+ = 0$ and $\phi_- = 0$
at the fixed points.

There are two types of radiative corrections to eq.\ (\ref{kktree}).
The first, called the bulk correction, comes from the fact that the theory
is not Lorentz invariant ($y$ is different from $x^\mu$) and therefore
the wavefunction renormalization $Z$ for the large dimensions is not equal
to $Z_5$, that for the fifth dimension. This correction, proportional to
$Z-Z_5$, occurs for loops that can sense the compactification ({\em i.e.},
if we flatten out the ED the loop definitely vanishes); 
it is in general small and
zero for fermions. While an exact expression is available in \cite{cms1}, 
these corrections are subdominant in the determination of the spectrum.

The second correction is called the orbifold correction. This is in general
log-divergent. In the presence of the $\sone/\ztwo$ orbifolding, the 
translational invariance along $y$ is lost. However, there is a discrete
$\ztwo$ symmetry, different from the earlier $\ztwo$, of $y\to y+\pi R$
which is still intact. The conservation of KK-number $n$ breaks down to
the conservation of KK-parity, defined as $(-1)^n$, which is a result of this
second $\ztwo$. Thus, (i) $n=1$ particles are to be produced in pairs
while $n=2$ particles can be singly produced; (ii) the lightest $n=1$
particle, often called the Lightest KK Particle (LKP), is the only
stable $n\not =0$ particle. This turns out to be an excellent CDM
candidate. In fact, this is completely analogous to the $\ztwo$ symmetry of the
underlying theory leading to a dark matter candidate for R-parity conserving
supersymmetry and the Little Higgs model with T-parity conservation. 

There are terms located on the fixed points that receive large log-divergent
contributions. One must include such terms to have a consistent
theory \footnote{The need for such terms can be understood as follows. One
can draw a 0-0-2 (these are KK numbers of the external legs) vertex at one-loop,
with $n=1$ particles flowing in the loop, which conserves KK number at all
vertices. However, this vertex is in general divergent. To cancel this 
divergence, one needs a divergent term itself in the original Lagrangian.}. 
To regulate these terms, one introduces a cut-off, $\Lambda$, upto
which the theory is said to be valid; hence the corrections come as 
$\log(\Lambda^2/\mu^2)$, where $\mu$, the regularisation scale, may be taken
to be $n/R$ for $n$-th level particles. $\Lambda$ should be at least of the
order of $R^{-1}$; phenomenologically, one takes $\Lambda R$ to be upto
50 or 100. The finite parts of these corrections are undetermined and remain
as free parameters of the theory; one may take, as a simplifying assumption,
the finite parts to vanish at the cutoff scale $\Lambda$ (that is one of the
principal assumptions of the minimal UED). Thus, the
orbifold corrections are of the form 
\be
\delta \propto \frac{1}{16\pi^2} f(g_i) \log\frac{\Lambda^2}{\mu^2}\,,
\ee
where $f(g_i)$ is a function depending upon the gauge couplings $g_i$
under which the field transforms nontrivially. For the exact expressions,
the reader is referred to \cite{cms1}. Note that just like any ED model,
UED is nonrenormalisable and should be treated as an effective theory valid
upto the cutoff scale.

%%------------------------------------------------------
\begin{figure}[htbp]
\vspace{10pt}
\centerline{%\hspace{-3.3mm}
\rotatebox{0}{\epsfxsize=8cm\epsfbox{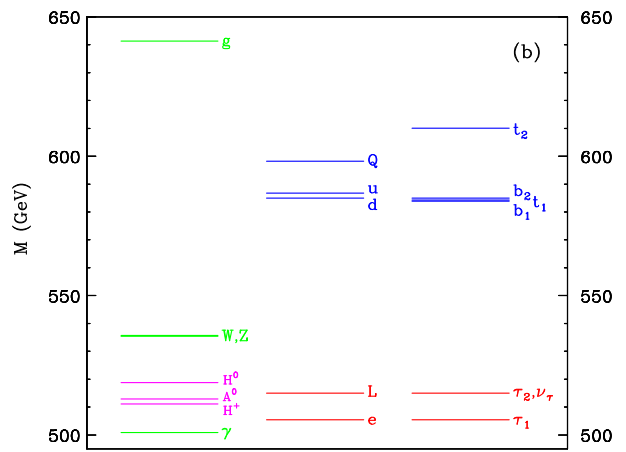}}}
\caption{$n=1$ levels for $R^{-1}=500$ GeV and $\Lambda R=20$. Taken from
\cite{cms1}.}
\label{fig1}
\end{figure}
%%-------------------------------------------------------

The following points are worth remembering:
\begin{itemize}
\item
The radiative corrections are positive. Therefore, the doublet quarks,
which are nonsinglet under all three SM gauge groups, receive the maximum
correction. The SU(2) singlet leptons (whose zero mode is right-chiral
but excitations are vectorial, with the $\ztwo$-even component being 
right-chiral) receive the minimum shift and are closest to the tree-level
mass as given in eq.\ (\ref{kktree}).
\item
While the vectorial mass terms $n/R$ and the radiative corrections do
not couple SU(2) singlet and doublet fermions, the Yukawa term does. This
is significant only for the top quark and hence the $n=1$ top quark masses
differ from their other charge $+2/3$ counterparts. 
\item
The gauge fields have five components. The first four, which are even under
$\ztwo$, appear as the excited gauge boson. The fifth component is a 
$\ztwo$-odd scalar. This mixes with the excitations of the Higgs doublet 
with the same quantum numbers. For example, the fifth component of $W^\pm$
mixes with the excitation of the charged Goldstone. Out of the two states,
one is eaten up by the $n=1$ gauge boson (the Higgs mechanism for the excited
level), while the other one remains in the physical spectrum. For $R^{-1}
\gg m_W,m_Z$, the physical particle is almost the excitation of the $n=0$
Higgs field. Thus, there are four $n=1$ scalars: $H^\pm$, $H^0$ (the excitation
of the SM Higgs boson), and $A^0$ (the excitation of the neutral CP-odd
Goldstone). The hierarchy $m_{H^\pm} < m_{A^0} < m_{H^0}$ is fixed, but
the spacing depends, among other factors, on the SM Higgs mass $m_h$.
There is one more term, $\overline{m_h^2}$, a soft term located
only at the fixed points, that affects {\em only} the scalar spectrum.
In the minimal UED, $\overline{m_h^2}$ is taken to be zero. More detailed
phenomenology of the scalar sector may be found in \cite{bbak2,bbak3} and
also in the talk \cite{biplob}. 
\item
At the $n=1$ level, $W^3$ and $B$ mix to give the physical states $Z_1$
and $\gamma_1$. However, the Weinberg angle $\theta_1$ is very small,
almost close to zero, so that $(W^3)_1 \approx Z_1$ and $B_1\approx
\gamma_1$. The latter is almost always the LKP. If one includes gravity,
the graviton becomes the LKP for $R^{-1}\leq 800$ GeV. Also, for very heavy
SM Higgs, the excited charged Higgs $H^\pm$ may turn out to be the LKP,
which in any way is cosmologically not viable. 
{\em Since $\gamma_1$ is neutral and weakly interacting, this
is an excellent CDM candidate.} As this is the end product of the cascade
of any $n=1$ state, the UED signal at the colliders, at least for the
production of $n=1$ states, is comparatively soft SM particles or jets
and {\em large missing energy} carried away by the LKP.
\end{itemize}

\section{Indirect Bounds on 1/R}

All SM particles have their corresponding towers. Thus, we expect 
finite radiative corrections from these heavy degrees of freedom on low-energy
observables, over and above the SM effects. We mention several such
observables and corresponding bounds on $R^{-1}$. Unlike collider signals,
these effects are not sensitive to the precise value of the cutoff scale
$\Lambda$. 

UED is a Minimal Flavour Violation (MFV) model. Such models are characterised
by the fact that there are no new CP violating phases apart from the one
present in the CKM matrix, which, in its $3\times 3$ form, is still unitary,
and no new FCNC operators apart from those occurring in the SM. MFV-type
models include two-Higgs doublet model, supergravity with small $\tan\beta$
and no new sources of FCNC (aligned quark and squark mass matrices), gauge
mediated SUSY, little Higgs with T-parity, {\em et cetera}.  

The new physics (NP) contribution to any MFV model is severely restricted.   
From the direct and indirect measurements of the sides and angles of the
Unitarity Triangle (UT), one can construct a so-called Universal Unitarity
Triangle, valid for all MFV-type models. The predictions are very close to
that of the SM; for example, $\sin(2\beta) = 0.735(0.732)\pm 0.049$, and
the tip of the UT $\bar{\rho}=0.174(0.187)\pm 0.068 (0.059)$, 
$\bar{\eta}=0.360(0.354)\pm 0.031(0.027)$, where the numbers in parenthesis
are those for the SM. Thus, just from CP-violating observables, it is
almost impossible to detect any evidence of any MFV-type model.

One can have excited gauge bosons, charged Higgs, and quarks, inside the
loop for the box diagram of $B^0$-$\overline{B}{}^0$ mixing. This 
contributes to the mass difference $\Delta M_d$ between the $B_d$ mass
eigenstates. While there are potentially infinite number of diagrams, 
coming from all $n$ upto infinity, it is enough to truncate the series
at, say, $\Lambda$. Also, for minimal UED, the result after such a 
truncation is finite and convergent. From the measured value of $\Delta M_d$,
the lower bound of $R^{-1}$ is about 250-300 GeV \cite{chk,buras}. A similar
consideration applies for the partial width of $Z$ to a $b\bar{b}$ final
state; the limit is roughly 300 GeV \cite{zbb}.   

A rather strong constraint comes from the radiative decay $B\to X_s+\gamma$
\cite{haisch}.
The experimental number for the branching fraction is $(3.55\pm 0.24 \pm 0.09
\pm 0.03)\times 10^{-6}$, where the first error is a combination of staistical
and systematic errors, the second one comes from the uncertainties in the
energy extrapolation, and the third one is due to the subtraction of 
$B\to X_d+\gamma$ events. The theoretical number, within the framework of
SM and at the NNLO level, is $(2.98\pm 0.26)\times 10^{-6}$. The uncertainties
include higher order perturbative effects, hadronic power corrections,
parametric dependences, and the uncertainty in the charm-quark mass. While
the numbers agree at about 1.1$\sigma$ level, any model, like minimal UED,
that decreases the branching ratio, is forced to be very tightly constrained.
For example, Haisch and Weiler \cite{haisch} found $1/R\geq 600$ GeV at
95\% CL. The calculation, however, takes the UED at LO for the matching of the
corresponding Wilson coefficients; two-loop effects with UED are yet to be
computed. The fact, together with the 99\% CL bound, still keeps open
a lower value of $R^{-1}\sim 400$ GeV, which may be interesting for a future
generation $e^+e^-$ machine.
  
The last indirect bound comes from the oblique parameters $S,T$ and $U$
\cite{acd,gogo}. The precision electroweak studies constrain $R^{-1}
\geq 600$ GeV for a light SM Higgs at about 115 GeV. However, this bound
is strongly sensitive on the SM Higs mass; for example, a 300 GeV Higgs 
will keep $R^{-1}\ge 400$ GeV open. The bound is also mildly sensitive
to the top mass.

\section{Collider Searches}

\subsection{Tevatron and LHC}

The role of colliders to investigate the possible nature of spacetime was
highlighted by Antoniadis \cite{earlycoll}. 
I will not say much about the Tevatron bounds as they have been superceded
by the indirect constraints. The LHC, however, is a different story. As we
will see later, $R^{-1}$ has a theoretically motivated upper bound of about
0.9-1 TeV, from dark matter abundance. Thus, the entire parameter space is 
accessible to LHC; if UED is there, at least in its minimal version, the
$n=1$ excitations are definitely going to be observed at the LHC
\cite{tev-lhc,tev-lhc2}. 

%%------------------------------------------------------
\begin{figure}[htbp]
\vspace{10pt}
\centerline{%\hspace{-3.3mm}
\rotatebox{0}{\epsfxsize=8cm\epsfbox{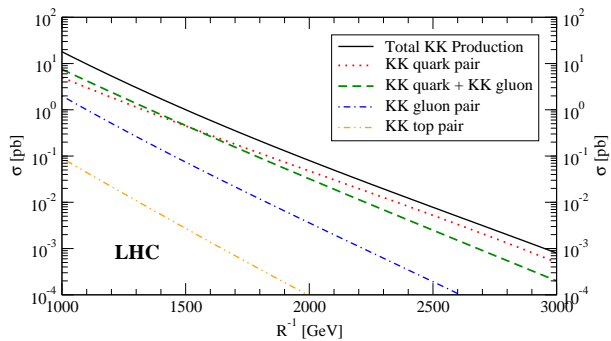}}}
\caption{Discovery potential at the LHC. From the first reference of
\cite{tev-lhc}.}
\label{fig2}
\end{figure}
%%-------------------------------------------------------

As the LKP is stable due to the underlying $\ztwo$ symmetry, this will
be the end product of the decay cascade of any $n=1$ particle that may 
be produced. The LKP escapes the detector, so all UED signals will
necessarily have large missing energy and transverse momentum. The 
near-degeneracy of the $n=1$ level forces the visible particles or jets to 
be comparatively soft. However, they should be visible unless coming from
a transition between very close levels; {\em e.g.}, the $\tau$s coming from
$H^+\to\nu_{\tau1} \tau_0$ are going to be very soft and will probably be below
the acceptance level of the detectors, so their detection
would be a challenge at the LHC. 

It has been pointed out in \cite{discrim} that the signals for UED, whether
they be soft leptons or jets accompanied by large missing $p_T$, can effectively
be faked by other NP models, most notably by supersymmetry with R-parity
conservation. For example, by replacing excited quarks and leptons by squarks
and sleptons, excited gauge bosons by corresponding gauginos, and LKP by
the lightest neutralino, the lightest SUSY particle (LSP), one can mimic
all signals of UED by SUSY. One of the possible discriminants at the LHC would
have been to determine the spin of the new particle, but in the unclean
environment of a hadron collider, this is a very tough job \cite{spin}.

%%------------------------------------------------------
\begin{figure}
\vspace{10pt}
\centerline{%\hspace{-3.3mm}
\rotatebox{-90}{\epsfxsize=8cm\epsfbox{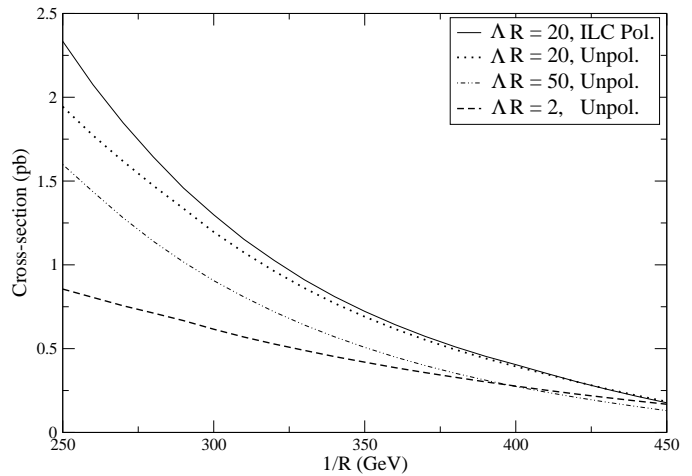}}}
\caption{Cross section versus $1/R$ for the process
$e^+e^-\to e^+e^- + $ missing energy. Plots are shown for unpolarised incident
beams with $\Lambda R =$ 2,20 and 50, and for `optimum' ILC polarisation (80\%
for $e^-$ and 50\% for $e^+$ beams) for $\Lambda R=$ 20. The lower and upper
energy cuts on the final state leptons are set at 0.5 and 20 GeV,
respectively. The angular cuts with respect to the beam axis are set at
$15^\circ$. From \cite{ilc1}.}
\label{fig3}
\end{figure}
%%------------------------------------------------------

Another option could be the production of $n=2$ gauge bosons, $\gamma_2$ and
$Z_2$, in $s$-channel. There are no analogues of these bosons in SUSY, and
so they may be thought of as the `smoking gun' of UED. However, this
question is subtle and we will return to it later. Here, one may note that
they can be singly produced as $s$-channel resonances, and hence do not
require more energy than that needed for the pair production of $n=1$ states.
Thus, if the collider is energetic enough to pair produce $n=1$ states, it
should produce $n=2$ resonances too. 

I have discussed earlier that there must be terms located at the fixed 
points $y=0,\pi R$ that go as $\log(\Lambda^2/\mu^2)$ for the UV completion
of the theory. They reduce the conservation of KK-number to the conservation
of KK-parity, $(-1)^n$. Thus, terms where an $n=2$ gauge boson is coupled to
two $n=0$ fermions are allowed; the coupling is small, suppressed by the
boundary-to-bulk ratio, but strong enough to produce the $n=2$ gauge bosons.
While the signal of $Z_2$ may be observed as a sharp bump in the dilepton
channel (these bosons, in turn, decay mostly into two $n=0$ fermions, so 
there is no missing energy), $\gamma_2$ will go unobserved. This is because
$\gamma_2$ couples almost entirely to $n=0$ quark pairs, and a dijet 
signal with no missing energy is definitely going to be swamped in the LHC
environment. 

This is where a next-generation $e^+e^-$ collider, like the International
Linear Collider (ILC), may come to our rescue. This is discussed in the
next subsection.

An important byproduct is the enhancement of the Higgs production rate at the
LHC. This goes mostly through gluon-gluon fusion, and with UED in the picture,
there is an excited top triangle that adds to the SM amplitude. For 
$m_H=150$ GeV and $R^{-1}=500 GeV$, this enhancement can be as high as 
80\% \cite{petriello}. 
 
\subsection{ILC}

The International Linear Collider is still in the blueprint stage.
This is going to be a single-pass linear collider, with two densely
packed beams of electron and positron hitting each other. The initial
centre-of-mass energy should be 500 GeV, an with sufficient motivation
(read discovery of new particles at the LHC) it can be upgraded to
1 TeV. 
ILC will provide a clean environment. This fact has motivated a number
of studies of UED at the ILC \cite{ilc0,ilc1,ilc2,ilc3,ilc4}. Let me just
highlight the major points.

\begin{itemize}
\item
ILC can discriminate between UED and SUSY. One just needs to look at the
differential decay distributions of the final-state electrons in $e^+e^-
\to e_1^+ e_1^-$ \cite{ilc1} or muons in $e^+e^-\to \mu_1^+\mu_1^-$
\cite{ilc2}. For
example, the distribution of $e^+e^-\to \mu^+\mu^-$ plus missing energy 
dips at $\theta=\pi/2$ for UED and peaks there for SUSY (the total number
of events will also be different). 

%%------------------------------------------------------
\begin{figure}[htbp]
\vspace{10pt}
\centerline{%\hspace{-3.3mm}
\rotatebox{0}{\epsfxsize=8cm\epsfbox{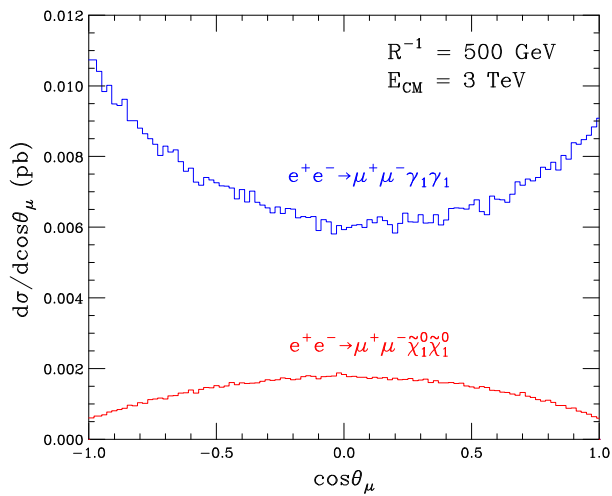}}}
\caption{Differential distribution for $e^+e^-\to \mu^+\mu^-$ plus
missing energy. From \cite{ilc2}.}
\label{fig4}
\end{figure}
%%-------------------------------------------------------

\item
ILC can act as a factory for production of the $n=2$ gauge bosons $\gamma_2$
and $Z_2$ \cite{ilc3}, in the same vein as LEP-1 which was a $Z$-factory. 
If ILC sits on one of these resonances, the production cross-section can be 
tens of picobarns. This, however, needs a prior knowledge of at least the
approximate positions of the peaks, which should be available from the LHC, 
and planning the ILC design accordingly. Unfortunately, ILC has predetermined
centre-of-mass energies that it will operate in, and it may be too late to
change that in view of the LHC data. 

\item
One can still salvage the situation and use the ILC to observe the narrow
$n=2$ resonances, even if they are away from the machine $\sqrt{s}$. This is
because the beam energy is degraded by the QED processes of initial state
radiation (ISR) and beamstrahlung; in both the cases, one or more photons is
radiated off the incoming particles so that the effective CM energy is less,
and there is a possibility that it will hit the resonance. This is nothing
but the phenomenon of radiative return, already observed in LEP-1.5. While the 
luminosities fall as we go away from the CM energy, the much enhanced 
cross-section at the resonance ensures that the signal is still visible
\cite{ilc4}. To observe such signals, ILC upgrade running at $\sqrt{s}=1
$ TeV will be required; a higher energy machine, like the proposed CLIC, 
will probe further in the parameter space. 

%%------------------------------------------------------
\begin{figure*}[htbp]
\vspace{10pt}
\centerline{%\hspace{-3.3mm}
\rotatebox{0}{\epsfxsize=15cm\epsfbox{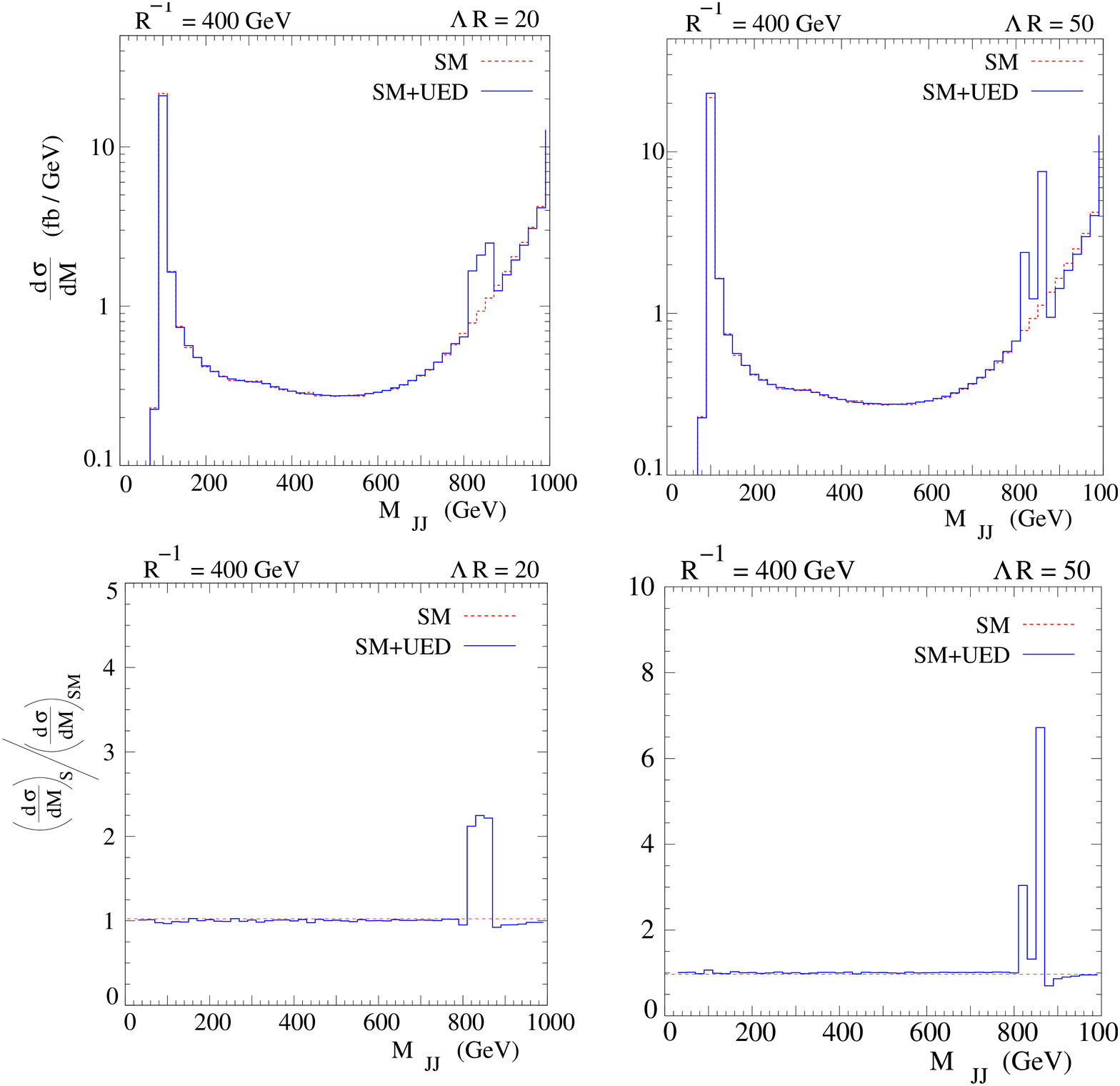}}}
\caption{Dijet mass distribution for two different values of $\Lambda$.
Also shown is the distribution normalised by the SM expectation. The
resolution is taken to be 20 GeV. 
%A more realistic resolution would be about 20
%GeV, with which the peaks for $\Lambda=20$ will not be resolved, but
%those for $\Lambda=50$ will still be. 
From \cite{ilc4}}
\label{fig4a}
\end{figure*}
%%-------------------------------------------------------

\item
While it has been established that UED can successfully be discriminated
from R-parity conserving SUSY at the ILC, the issue of discrimination from
other models should also be investigated. For example, R-parity violating
SUSY, with nonzero $\lambda$ and $\lambda'$-type couplings, can show 
$s$-channel peaks from sneutrino resonances, which are analogous to the 
$n=2$ resonances. Such peaks can be observed from RS gravitons, or one or
more extra $Z'$s. 

It has been shown in \cite{ilc4} that a simultaneous study of dijets and
4 lepton plus missing energy signals can act as a useful discriminator. 
The latter signal can come, in UED, from the process $e^+e^-\to Z_1Z_1$. 
Depending on the value of $\Lambda$, the dijet invariant mass distribution 
may show both $\gamma_2$ and $Z_2$ peaks, or they may be fused into one.  
Whatever the case might be, a sharp excess over the SM background is 
expected. The $4\ell$ plus missing energy signal also shows an excess.
Since all the leptons are expected to be soft, one can look for leptons,
none of whose $p_T$ should exceed 40 GeV. This reduces the SM background
to a negligible value; also, other competing models would give very different
distributions (for example, in a left-right symmetric model with extra
$W'$ and $Z'$, we do not expect soft leptons). 
Thus, a simultaneous study of both these signals in the
so-called signature space of the LHC should give an unambiguous map to
the multi-model parameter space.

\end{itemize}

\section{Astrophysical Bounds}

The best motivation of UED is, perhaps, a strong candidate for the cold
dark matter, viz., the LKP, which is mostly $B_1$, the excitation of the
hypercharge gauge boson. This is neutral and weakly interacting, and 
over a very large part of the parameter space, is the lightest $n=1$
particle (a small portion may have the $n=1$ charged Higgs as the LKP, 
but being charged, that is not a good CDM candidate). 

%%------------------------------------------------------
\begin{figure}[htbp]
\vspace{10pt}
\centerline{%\hspace{-3.3mm}
\rotatebox{0}{\epsfxsize=8cm\epsfbox{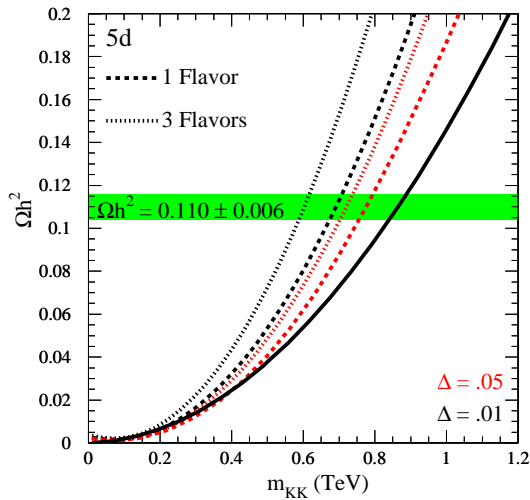}}}
\caption{The thermal relic abundance of the LKP without the effect
from any other KK species (solid line) and including the effects of
KK leptons 5\% and 1\% heavier than the LKP (dashed and dotted lines).
The horizontal band gives the measured DM abundance \cite{wmap}.
From the third reference of \cite{dark1}.}
\label{fig5}
\end{figure}
%%-------------------------------------------------------

The KK excitations, and hence the LKP, were freely produced and annihilated
in the early universe, where $T\gg R^{-1}$. As the universe expands and cools,
the LKP decouples and freezes out, and forms a thermal relic. The number 
density of the LKP can be estimated using the Boltzmann equation, and 
the LKP self-annihilation rate, whose cross-section is roughly $95 g_1^4/
324\pi m_{\gamma_1}^2$, where $g_1$ is the $U(1)_Y$ coupling constant
\cite{dark1,dark2}. The most accurate result is the one quoted in the last reference of \cite{dark1}.
The relic abundance has been measured by WMAP to be 
$\Omega h^2 = 0.110\pm 0.006$ \cite{wmap}, which translates roughly to
a window between 850 and 900 GeV for the LKP. If the CDM is not entirely
due to the LKP, the lower limit may be relaxed, but the upper limit should
be about $R^{-1} \leq 900$ GeV. This is the result that makes us confident
about the discovery of UED, provided it is the path Nature takes, at the
LHC. 

There may be other species of $n=1$ particles close to the LKP. The
best candidates for this are the excited leptons and neutrinos. If they
are sufficiently close, their effects on co-annihilation should be taken into
account. The closer they are, the tighter is the upper bound on $R^{-1}$.
This is shown in figure \ref{fig5}. All in all, one can say that the parameter
space of UED should conservatively be in the range 400 GeV$<R^{-1}<$ 900 GeV.
For a more detailed review, I would suggest ref.\ \cite{dark2}.

\subsection{Gravity in UED}

If one includes gravity in UED, place must be provided for the KK excitations
of gravitons, whose $n$-th level should be simply at $n/R$ \cite{gravity}.
The $n=1$ graviton immediately becomes the LKP upto $R^{-1}=810$ GeV, 
above which
the radiative corrections on $\gamma_1$ push it below $G_1$. However, the
graviton LKP scenario is not viable. In the graviton LKP region, 
$\gamma_1$ decays to $G_1$ and a photon. The contribution of this process
to the diffuse photon flux is much above the experimental limit, ruling
the graviton scenario out. If one includes gravity, there is a small patch
of the parameter space, with $R^{-1}>810$ GeV, where $\gamma_1$ is the LKP
and still does not violate the dark matter overproduction constraint. 
The $n=1$ excitations should be observable at the LHC but the ILC, even
with the upgraded $\sqrt{s}=1$ TeV option, will draw a blank. The model
also becomes rather fine-tuned and loses much of its charm as far as the
collider search is concerned.

\section{6-d UED}

The 6-d UED, where there are two extra dimensions accessible to the SM
particles, has some strong phenomenological motivations. It is well-known 
that the proton stability is a problematic issue in any extra dimensional
scenario, because of the new higher dimensional operators that may lead 
to an unacceptably quick decay of the proton. In 6-d UED, the global
symmetries of the theory prevents all proton decay operators less than
dimension 9 \cite{6dproton}. For example, the decay $p\to e^- \pi^+\pi^+
\nu\nu$ has a lifetime of $10^{35}$ years for $R^{-1}=500$ GeV and 
$\Lambda R=5$, and scales as $(R^{-1}/500)^{12}$ and $(\Lambda R/5)^
{22}$! 

It was shown in \cite{6d3gen} that an $SU(2)_L$ global gauge anomaly
exists unless the number of $\ztwo$-even doublets differ from that of
$\ztwo$-odd doublets by an integral multiple of six. For each generation, 
this difference is either 2 or 4; thus, one needs three generations (or
a multiple of three) for the anomaly cancellation.

To get chiral fermions, the 6-d UED needs to be compactified on a
chiral square \cite{6dforma}. This is a square with adjacent sides identified.
Each SM particle has excitations specified by two positive integers
$(j,k)$, so that the mass of the $(j,k)$-th excitation, at the tree-level,
is given by
\be
m_{j,k}^2 = m_0^2 + \frac{j^2+k^2}{R^2}\,.
\ee

The scalar sector of 6-d UED is richer than its 5-d counterpart. Each gauge
field has 6 degrees of freedom; hence, there is an extra adjoint scalar in
the spectrum for each gauge field. The scalar adjoint $B_{1,0}$ is the LKP
and turns out to be a good dark matter candidate. For more phenomenological
analysis, the reader may look at \cite{6dpheno}, and also in the talk
\cite{kirtiman}.

\begin{acknowledgments}
I thank the organisers of PWED, in particular Sayan Kar, for such a 
stimulating workshop. This work was supported by the research grants 
SR/S2/HEP-15/2003 of DST, Govt.\ of India, and 2007/37/9/BRNS
of DAE, Govt.\ of India.
\end{acknowledgments}

\end{document}